\documentclass[runningheads]{llncs}

\usepackage[
    lambda, advantage, operators, sets, adversary, landau, probability, notions,
    logic, ff, mm, primitives, events, complexity, asymptotics, keys
]{cryptocode}

\usepackage{mathtools}
\usepackage{amsmath}
\usepackage{amssymb}

\usepackage{graphicx}
\usepackage[position=bottom]{subfig}
\usepackage{tikz}
\usepackage{tikz-qtree}
\usetikzlibrary{shapes.misc,positioning,arrows,snakes}

\usepackage{color}
\usepackage{colortbl}
\definecolor{myteal}{HTML}{009999}
\definecolor{mylime}{HTML}{809933}
\definecolor{mypurple}{HTML}{993380}
\definecolor{myred}{HTML}{E04644}
\definecolor{mydgreen}{HTML}{008000}
\definecolor{mydblue}{HTML}{2809B2}

\usepackage{hyperref}
\hypersetup{
  colorlinks = true,                              
  urlcolor   = mydblue,                               
  linkcolor  = mydblue,                                     
  citecolor  = mydgreen,                                         
}

\title{%
  Verifiable Light-Weight Monitoring for Certificate Transparency Logs
}
\titlerunning{%
  Verifiable Light-Weight Monitoring for Certificate Transparency Logs
}

\author{%
  Rasmus~Dahlberg \and
  Tobias~Pulls%
}

\authorrunning{%
  R.\ Dahlberg \and T.\ Pulls
}

\institute{%
  Dept.\ of Mathematics and Computer Science,
  Karlstad University, Sweden
  \email{first.last@kau.se}
}

\begin{document}
  \maketitle
  \begin{abstract}
    Trust in publicly verifiable Certificate Transparency (CT) logs is reduced
through
  cryptography,
  gossip,
  auditing, and
  monitoring.
The role of a monitor is to observe each and every log entry, looking for
suspicious certificates that interest the entity running the monitor.
While anyone can run a monitor, it requires
  continuous operation and
  copies of the logs to be inspected.
This has lead to the emergence of monitoring as-a-service:
  a trusted third-party runs the monitor and provides registered subjects with
  selective certificate notifications.
We present a CT/bis extension for verifiable \emph{light-weight monitoring} that
enables subjects to verify the correctness of such certificate notifications,
making it easier to distribute and reduce the trust which is otherwise placed in
these monitors. Our extension
supports verifiable monitoring of wild-card domains and piggybacks on CT's
existing gossip-audit security model.

  \end{abstract}

  \section{Introduction}
Certificate Transparency (CT)~\cite{ct} is an experimental standard that
enhances the public-key infrastructure by adding transparency for certificates
that are issued by Certificate Authorities (CAs). The idea is to mandate
that every certificate must be publicly logged in an append-only tamper-evident
data structure~\cite{history-tree}, such that anyone can observe what has been
issued for whom. This means that a subject can determine for herself if anything
is mis-issued by downloading all certificates; so called \emph{self-monitoring}.
An alternative monitoring approach is to rely on a trusted third-party that
\emph{notifies} the subject if relevant certificates are ever found. Given that
self-monitoring involves set-up, continuous operation, and exhaustive
communication effort, the concept of subscribing for monitoring
\emph{as-a-service} is simpler for the subject. This model is already prevalent
in the wild, and is provided both by CAs and industry vendors---see for example
SSLMate's \emph{Cert Spotter}\footnote{%
  \url{https://sslmate.com/certspotter/}, accessed 2018-09-15.
} or Facebook's monitoring tool\footnote{%
  \url{https://developers.facebook.com/tools/ct/}, accessed 2018-09-15.
}. Third-party monitors can also offer related services, such as searching for
certificates interactively or inspecting other log properties. The former is
provided by Facebook and Comodo's \url{crt.sh}; the latter by Graham Edgecombe's
CT monitoring tool\footnote{%
  \url{https://ct.grahamedgecombe.com/}, accessed 2018-09-15.
}.

It would be an unfortunate short-coming if CT did not change the status quo of
centralized trust by forcing subjects who cannot operate a self-monitor to trust
certificate notifications that are provided by a third-party monitor. While it
is true that a subject could subscribe to a large number of monitors to reduce
this trust, it is overall cumbersome and does not scale well beyond a handful
of notifying monitors (should they exist). To this end, we suggest a CT/bis
extension for verifiable Light-Weight Monitoring (LWM) that makes it easier to
distribute the trust which is otherwise placed in these monitors by decoupling
the notifier from the full-audit function of inspecting all certificates. Our
idea is best described in terms of a self-monitor that polls for new updates,
but as opposed to processing all certificates we can filter on wild-card
prefixes such as \texttt{*.example.com} in a verifiable manner.
LWM relies on the ability to define a new Signed Tree Head (STH) extension, and
thus a CT/bis compliant log is necessary~\cite{ct/bis}. At the time of writing
CT/bis have yet to be published as an IETF standard. We are not aware of any log
that deploys a drafted version.

As a brief overview, each batch of newly included certificates are grouped as a
static Merkle tree in LWM. The resulting snapshot (also know as a fingerprint or
a root hash) is then incorporated into the corresponding STH as an extension.
An LWM subject receives one verifiable certificate notification per log update
from an untrusted \emph{notifier}
  (who could be the log, a monitor, or anyone else),
and this notification is based on the smaller static Merkle tree rather than the
complete log. This is because monitoring as-a-service is mainly about
identifying newly included certificates. Moreover, we can order each static
Merkle tree so that verifiable wild-card filtering is possible. For security we
rely on at least one entity to verify that each snapshot is correct---which is
a general monitoring function that is independent of the subjects using LWM---%
as well as a gossip protocol that detects split-views~\cite{sthnp}. Since our
extension is part of an STH, we piggyback on any gossip-like protocol that deals
with the exchange and/or distribution of (verified) STHs~%
  \cite{ctga,ietf-gossip,sth-push,cosi}.
Our contributions are as follows:
\begin{itemize}
  \item The design of a backwards-compatible CT/bis extension for light-weight
    monitoring of wild-card prefixes such as \texttt{*.example.com}
      (Section~\ref{sec:lwm}).
  \item A security sketch showing that an attacker cannot omit a certificate
    notification without being detected, relying on standard cryptographic
    assumptions and piggybacking on the proposed gossip-audit models of CT
      (Section~\ref{sec:eval:security}).
  \item An open-source proof-of-concept implementation written in Go, as well
    as a performance evaluation that considers computation time and bandwidth
    requirements (Section~\ref{sec:eval:perf}). In particular we find that the
    overhead during tree head construction is small in comparison to a sound STH
    frequency of one hour; a notifier can easily notify 288~M subjects in a
    verifiable manner for Google's Icarus log on a single core and a 1~Gbps
    connection; and a subject receives about 24~Kb of proofs per day and log
    which is verified in negligible time (the order of $\mu$s for the common
    case of non-membership, and seconds in the extreme case of verifying
    membership for \emph{an entire top-level domain}).
\end{itemize}

Background on Merkle trees and CT is provided in Section~\ref{sec:background}.
Related work is discussed in Section~\ref{sec:eval:related}.
Conclusions are presented in Section~\ref{sec:conclusion}.

  \section{Background} \label{sec:background}
Suppose that a trusted content provider would like to outsource its operation to
an untrusted third-party. This is often referred to as the three-party setting,
in which a trusted source maintains an authenticated data structure through a
responder that answers client queries on the source's behalf~\cite{ads}.
The data structure is authenticated in the sense that every answer is
accompanied by a cryptographic proof that can be verified for correctness by
only trusting the source.
While there are many settings and flavors of authenticated data
structures~\cite{history-tree,pad,accumulator}, our scope is narrowed down to CT
which builds upon Merkle trees.

\subsection{Merkle Trees} \label{sec:background:mt}
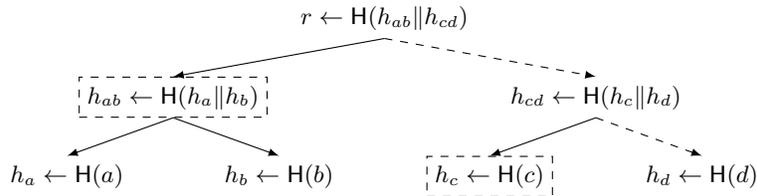
\begin{figure}
  \centering
  \begin{tikzpicture}[
  sibling distance=32pt,
  -latex,
  apnode/.style = {
    draw=black,
    dashed,
  },
  ap/.style = {
    draw=black,
    dashed,
  },
]
  \Tree [
    .$r\gets\hash(h_{ab}\concat h_{cd})$ [
      .\node[apnode]{$h_{ab}\gets\hash(h_a\concat h_b)$}; [
        .$h_a\gets\hash(a)$
      ] [
        .$h_b\gets\hash(b)$
      ]
    ] \edge[ap]; [
      .$h_{cd}\gets\hash(h_c\concat h_d)$ [
        .\node[apnode]{$h_c\gets\hash(c)$};
      ] \edge[ap]; [
        .$h_d\gets\hash(d)$
      ]
    ]
  ]   
\end{tikzpicture}
  \caption{%
    Merkle tree containing four values $a$--$d$. The dashed arrows show the
    traversal used to generate an audit path for the right-most leaf (dashed
    nodes).
  }
  \label{fig:mt}
\end{figure}

The seminal work by Merkle~\cite{mt} proposed a \emph{static}  binary tree where
each leaf stores the hash of a value and every interior node hashes its children
  (Figure~\ref{fig:mt}).
The root hash serves as a succinct snapshot of the tree's structure and content,
and by revealing a logarithmic number of hashes it can be reconstructed to prove
whether a value is stored in a leaf. These hashes compose an audit path for
a value, and it is obtained by taking every sibling hash while traversing the
tree from the root down towards the leaf being authenticated. An audit path is
verified by reversing the traversal used during generation, first reconstructing
the leaf hash and then every interior node recursively
  (using the provided sibling hashes) 
until finally reaching the root. Given a collision resistant hash function,
an audit path proves that a given leaf contains a value iff the reconstructed
root hash is known to be authentic. For example, the trusted source might sign
it.

While non-membership of a value can be proven by providing the entire data
structure, this is generally too inefficient since it requires linear space and
time. A better approach is to structure the tree such that the node which should
contain a value is known if it exists. This property is often discussed in
relation to certificate revocation:
  as opposed to downloading a list of serial numbers that represent the set of
    revoked certificates,
  each leaf in a static Merkle tree could (for example) contain an interval
    $[a, b)$ where $a$ is revoked and the open interval $(a,b)$
    current~\cite{crt}.
Given a serial number $x$, an audit path can be generated in logarithmic space
and time for the leaf where $x \in [a,b)$ to prove (non-)membership. Similar
constructions that are \emph{dynamic} support updates more efficiently~%
\cite{pad,vds,coniks}.

\subsection{Certificate Transparency} \label{sec:bac:ct}
The CA ecosystem involves hundreds of trusted third-parties that issue TLS
certificates~\cite{ca-ecosystem}. Once in a while \emph{somebody} gets this process
wrong, and as a result a fraudulent identity-to-key binding may be issued for
\emph{any} subject~\cite{enisa}.
It is important to detect such incidents because mis-issued certificates can
be used to intercept TLS connections. However, detection is hard unless the
subjects \emph{who can distinguish between anything benign and fraudulent}
get a concise view of the certificates that are being served to the clients.
By requiring that every CA-issued certificate must be disclosed in a public
and append-only log, CT layers on-top of the error-prone CA ecosystem to provide
such a view:
  in theory anyone can inspect a log and determine for herself if a certificate
  is mis-issued~\cite{ct}.

It would be counter-intuitive to `solve' blind trust in CAs by suggesting that
everybody should trust a log. Therefore, CT is designed such that the log can
be distrusted based on two components:
  a dynamic append-only Merkle tree that supports verifiable membership and
    consistency queries~\cite{history-tree}, as well as
  a gossip protocol that detects split-views~%
    \cite{sthnp,ietf-gossip}.
We already introduced the principles of membership proofs in Section~%
\ref{sec:background:mt}, and consistency proofs are similar in that a
logarithmic number of hashes are revealed to prove two snapshots consistent.
In other words, anyone can verify that a certificate is included in the log
without fully downloading it, and whatever was in the log before still remains
unmodified. Unlike the three-party setting, gossip is needed because there is no
trusted source that signs-off the authenticated data structure:
  consistency and inclusion proofs have limited value if everybody observes
  different (but valid) versions of the log.

\subsubsection{Terminology, policy parameters and status quo.}
A new STH---recall that this is short for Signed Tree Head---is issued by the
log at least every Maximum Merge Delay (MMD) and no faster than allowed by an
STH frequency~\cite{ct/bis}. An MMD is the
longest time until a certificate must be included in the log after promising to
include it. This promise is referred to as a Signed Certificate Timestamp (SCT).
An STH frequency is relative to the MMD, and limits the number of STHs that can
be issued. These parameters (among others) are defined in a log's policy, and if
a violation is detected there are non-repudiable proofs of log misbehavior that
can be presented. For example, show
  an SCT that is not included after an MMD,
  too many STHs during the period of an MMD, or
  two STHs that are part of two inconsistent versions of the log.
In other words, rather than being a trusted source a log signs statements to be
held accountable.

Ideally we would have all of these components in place at once: anyone that
interacts with a log audits it for correctness based on partial information
  (SCTs, STHs, served certificates, and proofs),
subjects monitor the logs for newly included certificates to check that they are
free from mis-issuance (full download), and a gossip protocol detects or deters
logs from presenting split-views. This is not the case in practice, mainly
because CT is being deployed incrementally~\cite{sth-push}
but also because the cost and complexity of self-monitoring is relatively high.
For example,
a subject that wants rapid detection of mis-issuance needs continuous operation
and full downloads of the logs. It appears that the barrier towards self-%
monitoring have lead to the emergence of monitoring as-a-service, where a
trusted third-party monitors the logs on a subject's behalf by selectively
notifying her of relevant certificates, e.g., mail the operator of
$\mathsf{example.com}$ if $\mathsf{*.example.com}$ certificates are ever found.
Third-party monitoring is convenient for logs too because it reduces the
bandwidth required to serve many subjects. However, for CT it is an unintuitive
concept given that it requires blind trust.

  \section{Light-Weight Monitoring} \label{sec:lwm}
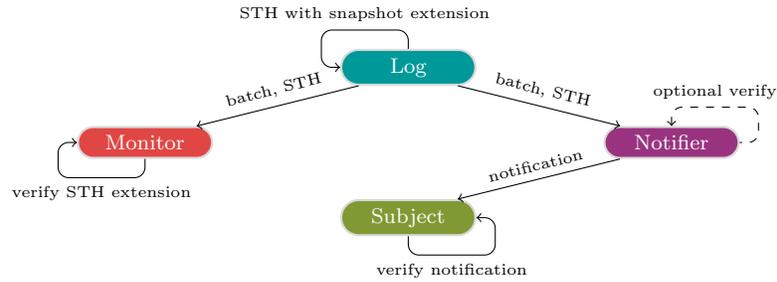
\begin{figure}
  \centering
  \begin{tikzpicture}[
  -latex,
  rrs/.style = {
    draw = gray!30,
    thick,
    rounded rectangle,
    fill = white,
    minimum width = 2cm,
    font = \fontsize{8}{8}\selectfont,
    text = white,
  },
  ls/.style = {
    font=\fontsize{6}{6}\selectfont,
  },
]
\draw (0, 1) node[rrs, fill=myteal] (Log) {Log};
\draw (0, -1) node[rrs, fill=mylime] (Subject) {Subject};
\draw (3.5, 0) node[rrs, fill=mypurple] (Endpoint) {Notifier};
\draw (-3.5, 0) node[rrs, fill=myred] (Monitor) {Monitor};

\path [draw, ->, rounded corners]
  (Log.north) |-
  ($ (Log.north) + (Log.west) - (Log) + (-0.25, 0.25) $)
    node[ls, above, pos=0.75]{
      STH with snapshot extension
    } |-
  (Log.west);

\path [draw, ->, rounded corners]
  (Monitor.south) |-
  ($ (Monitor.south) + (Monitor.west) - (Monitor) + (-0.25, -0.25) $)
    node[ls, below, pos=0.75]{
      verify STH extension
    } |-
  (Monitor.west);

\path [draw, ->, rounded corners]
  (Subject.south) |-
  ($ (Subject.south) + (Subject.east) - (Subject) + (0.25, -0.25) $)
    node[ls, below, pos=0.75]{
      verify notification
    } |-
  (Subject.east);

\path [draw, <-, dashed, rounded corners]
  (Endpoint.north) |-
  ($ (Endpoint.east) + (Endpoint.north) - (Endpoint) + (0.25, 0.25) $)
    node[ls, above, pos=0.75]{
      optional verify
    } |-
  (Endpoint.east);

\draw [->]
  (Log.south east) --
    node[ls, sloped, anchor=center, above]{%
      batch, STH
    }
  (Endpoint.north west);

\draw [->]
  (Endpoint.south west) --
    node[ls, sloped, anchor=center, above]{%
      notification
    }
  (Subject.north east);

\path [draw, ->]
  (Log.south west) --
    node[ls, sloped, anchor=center, above]{%
      batch, STH
    }
  (Monitor.north east);
\end{tikzpicture}
  \caption{%
    An overview of LWM. In addition to normal operation, a log creates an
    additional (smaller) Merkle tree that supports wild-card (non-)membership
    queries. The resulting snapshot is signed as part of an STH extension that
    can be verified by any monitor that downloads the corresponding batch. A
    subject receives one verifiable certificate notification per STH from an
    untrusted notifier.
  }
  \label{fig:idea}
\end{figure}

To reduce the trust which is placed in today's third-party monitors,
the idea of LWM is to lower the barrier towards self-monitoring. As shown in
Figure~\ref{fig:idea}, an untrusted notifier provides a subject with
efficient\footnote{%
  Efficient iff less than a linear number of log entries are received per log
	update.
} certificate notifications that can be cryptographically verified: each batch
of certificates is represented by an additional Merkle tree that supports
wild-card (non-)membership queries
  (described further in Section~\ref{sec:lwm:wildcard}),
and the resulting snapshot is signed by the log as part of an STH extension.
As such, a subject can deal only with those certificates that are relevant,
relying on wild-card proofs to verify correctness and completeness:
  said certificates are included and nothing is being omitted.
Anyone can check that an LWM snapshot is correct by inspecting the corresponding
batch of certificates. Notably this is \emph{a general monitoring function},
rather than a \emph{selective notification component} which is verifiable in
LWM. This decoupling allows anyone to be a notifier, including logs and
monitors that a subject distrust.

\subsection{Authenticated Wild-Card Queries} \label{sec:lwm:wildcard}
\begin{figure}
  \centering
  \begin{tikzpicture}[
  sibling distance=6pt,
  level distance=100pt,
  -latex,
  grow=left,
]
  \Tree [
    .$r\gets\hash(h_{01}\concat h_{23})$ [
      .$h_{01}\gets\hash(h_0\concat h_1)$ [
        .$h_0\gets\hash(\mathsf{gro.elpmaxe})$
      ] [
        .$h_1\gets\hash(\mathsf{moc.elpmaxe})$
      ]
    ] [
      .$h_{23}\gets\hash(h_2\concat h_3)$ [
        .$h_2\gets\hash(\mathsf{moc.elpmaxe.bus})$
      ] [
        .$h_3\gets\hash(\mathsf{ten.elpmaxe})$
      ]
    ]
  ]   
\end{tikzpicture}
  \caption{%
    Merkle tree where the leaves are ordered on reversed subject names.
  }
  \label{fig:wildcard}
\end{figure}
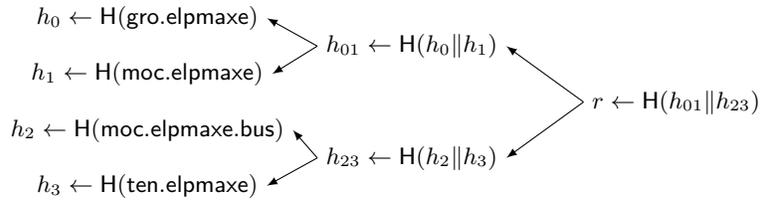

Thus far we only discussed Merkle trees in terms of verifying whether a single
value is a (non-)member:
  membership is proven by presenting an audit path down to the leaf in question,
    while
  non-membership requires a lexicographical ordering that allows a verifier
    to conclude that a value is absent unless provided in a particular location.
The latter concept naturally extends to \emph{prefix wild-card queries}---such
as $\mathsf{*.example.com}$ and $\mathsf{*.sub.example.com}$---by finding a
suitable ordering function $\Omega$ which ensures that related leaves are
grouped together as a consecutive range. We found that this requirement is
satisfied by sorting on reversed subject names:
  suppose that we have a batch of certificates
    $\mathsf{example.com}$,
    $\mathsf{example.org}$,
    $\mathsf{example.net}$, and
    $\mathsf{sub.example.com}$.
After applying $\Omega$ we get the static Merkle tree in Figure~%
\ref{fig:wildcard}. A prefix wild-card proof is constructed by finding the
relevant range in question, generating an audit path for the
leaves that are right outside of the range~\cite{range-mt}. Such a proof is
verified by checking that
  (i) $\Omega$ indicates that the left (right) end is less (larger) than the
    queried prefix,
 (ii) the leaves are ordered as dictated by $\Omega$, and
(iii) the recomputed root hash is valid.

The exact details of reconstructing the root hash is a bit tedious because there
are several corner cases. For example, either or both of the two audit paths may
be empty depending on batch size (${\leq}1$) and location of the relevant range
(left/right-most side). Therefore, we omit the details and
focus on the concept:
  given two audit paths and a sequence of data items ordered by $\Omega$ that
    includes the left leaf, matching range, and right leaf, repeatedly
    reconstruct interior nodes to the largest extent possible and then use the
    sibling hash which is furthest from the root to continue.
For example, consider a proof for $\mathsf{*sub.example.com}$ in Figure~%
\ref{fig:wildcard}: it is composed of
  (i) the left leaf data and its audit path $h_0,h_{23}$ on index 1,
 (ii) the right leaf data and its audit path $h_2,h_{01}$ on index 3, and
(iii) the matching range itself which is a single certificate.
After verifying $\Omega$~order, recompute the root hash $r'$ and check if
it matches an authentic root $r$ as follows:
\begin{enumerate}
  \item Compute leaf hashes $h_1'$, $h_2'$, and $h_3'$ from the provided data.
    Next, compute the interior node $h_{23}' \gets \hash(h_2'\concat h_3')$.
    Because no additional interior node can be computed without a sibling hash,
    consider $h_0$ in the left audit path.
  \item Compute the interior node $h_{01}' \gets \hash(h_0\concat h_1')$, then
    finally $r' \gets \hash(h_{01}'\concat h_{23}')$.\footnote{%
      Two audit paths may contain redundancy, but we ignored this favouring
      simplicity.
    }
\end{enumerate}

Given an $\Omega$~ordered list of certificates it is trivial to locate where
a subject's wild-card matches are:
  binary search to find the index of an exact match (if any), then up to
  $t$ matches follow in order.
This is not the only way to find the right range and matches. For example, a
radix tree could be used with the main difference being $\bigO{t+\log n}$
against $\bigO{t+k}$ complexity for a batch of size $n$, a wild-card string of
length $k$, and $t$ matches. Since the complexity of generating two audit
paths is $\bigO{\log n}$ for any number of matches, the final space and time
complexity for a wild-card structure based on an ordered list is
$\bigO{t+\log n}$.

\subsection{Notifier} \label{sec:lwm:notifier}
A notifier must obtain every STH to generate wild-card proofs that can be traced
back to the log. Albeit error-prone in case of network issues, the simplest
way to go about this is to poll the log's get-STH endpoint \emph{frequently
enough}.\footnote{
  It would be better if logs supported verifiable and historical get-STH
  queries.
}
Once an updated is spotted every new certificate is downloaded and the wild-card
structure is reconstructed. A subject receives her verifiable certificate
notifications from the notifier via a push (`monitoring as-a-service') or pull
(`self-monitoring') model. For example, emails could be delivered after every
update or in daily digests. Another option is to support queries like
  ``what's new since STH~$x$''.

A subject can verify that a certificate notification is fresh by inspecting the
STH timestamp. However, it is hard to detect missing certificate notifications
unless every STH trivially follows from the previous one. While there are
several methods to achieve this---%
  for example using indices (Section~\ref{sec:lwm:instantiation}) or
  hash chains~\cite{coniks}---%
the log must always sign a snapshot per STH using an extension.

\subsection{Instantiation Example} \label{sec:lwm:instantiation}
Instantiating LWM depends upon the ability to support an STH extension. In the
latest version of CT, this takes the form of a sorted list of
key-value pairs where the key is unique and the value an opaque byte array~%
\cite{ct/bis}. We could reserve the keywords
  \emph{lwm} for snapshots and
  \emph{index} for monotonically increasing counters.\footnote{%
		Instead of an index to detect missing notifications (STHs), a log could
    announce STHs as part of a verifiable get-STH endpoint. See the sketch of
    Nordberg:
			\url{https://web.archive.org/web/20170806160119/https://mailarchive.ietf.org/arch/msg/trans/JbFiwO90PjcYzXrEgh-Y7bFG5Fw}, accessed 2018-09-16.
} Besides an LWM-compliant log, an untrusted notifier must support pushed or
pulled certificate notifications that are verifiable by tracking the most recent
or every wild-card structure. Examples of likely notifiers include
  logs (who benefit from the reduced bandwidth) and
  monitors (who could market increased transparency)
that already process all certificates regardless of LWM.

  \section{Evaluation} \label{sec:evaluation}
First we discuss assumptions and sketch on relevant security properties for LWM.
Next, we examine performance properties of our open-source proof-of-concept
implementation experimentally and reason about bandwidth overhead in theory.
Finally, we present differences and similarities between LWM and related work.

\subsection{Assumptions and Security Notions} \label{sec:eval:security}
The primary threat is a computationally bound attacker that attempts to forge or omit a
certificate notification without being detected.
We rely on standard cryptographic assumptions, namely an unforgeable digital
signature scheme and a collision resistant hash function $\hash$ with
$2\secpar$-bit output for a security parameter~$\secpar$.
The former means that an LWM snapshot must originate from the (untrusted) log in
question. While an incorrect snapshot could be created intentionally to hide a
mis-issued certificate, it would be detected if at least one honest monitor
exists because our STH extension piggybacks on the gossip-audit model of CT
(that we assume is secure).\footnote{%
  Suppose that witness cosigning is used~\cite{cosi}. Then we rely on at least
  one witness to verify our extension. Or, suppose that STH pollination is
  used~\cite{ietf-gossip}. Then we rely on the most recent window of STHs to
  reach a monitor that verifies our extension.
}
A subject can further detect missing notifications by checking the STH index for
monotonic increases and the STH timestamp for freshness. Thus, given secure
audit paths and correct verification checks as described in Section~%
\ref{sec:lwm:wildcard}, no certificate notification can be forged or omitted.
Our cryptographic assumptions ensure that every leaf is fixed by a secure
audit path as in CT, i.e., a leaf hash with value $v$ is encoded as
$\hash(0x00 \concat v$) and an interior hash with children $L,R$ as
$\hash(0x01 \concat L \concat R)$~\cite{history-tree,ct}. To exclude any
unnecessary data on the ends of a range, the value $v$ is a subject name
concatenated with a hashed list of associated certificates in LWM (subject
names suffice to verify $\Omega$~order).

CT makes no attempt to offer security in the multi-instance setting~\cite{katz}.
Here, an attacker that targets many different Merkle trees in parallel should
gain no advantage while trying to forge \emph{any} valid (non-)membership
proof. By design there will be many different wild-card Merkle trees in LWM, and
so the (strictly stronger) multi-instance setting is reasonable. We can
provide full bit-security in this setting by ensuring that no node's pre-image
is valid across different trees by incorporating a unique tree-wide
constant $c_t$ in leaf and empty hashes \emph{per batch}, e.g.,
$c_t \sample \set{0,1}^\secpar$. Melera \emph{et~al.}~\cite{coniks}
describe this in detail while also ensuring that no node's pre-image is valid
across different locations within a Merkle tree.

In an ecosystem where CT is being deployed incrementally without gossip, the
benefit of LWM is that a subject who subscribes for certificate notifications
can trust the log only (as opposed to \emph{also} trusting the notifier).
Therefore, today's trust in third-party monitoring services can be reduced
significantly. A log must also present a split-view or an invalid snapshot to
deceive a subject with false notifications. As such, subjects accumulate binding
evidence of log misbehavior that can be audited sometime in the future if
suspicion towards a log is raised. Long-term the benefit of LWM is that it is
easier to distribute the trust which is placed in third-party monitors, i.e.,
anyone who processes a (small in comparison to the entire log) batch of
certificates can full-audit it without being a notifier.

\subsection{Implementation and Performance} \label{sec:eval:perf}
We implemented multi-instance secure LWM in less than 400 lines of Go.%
  \footnote{%
    Open source implementation available at \url{https://github.com/rgdd/lwm}.
  }
Our wild-card structure uses an existing implementation of a radix tree to find
leaf indices and data. To minimize proof-generation times, all hashes are
cached in an in-memory Merkle tree which uses SHA-256. We benchmarked snapshot
creation, proof generation, and proof verification times on a single core as the
batch size increases from 1024--689,245~certificates using 
  Go's built-in benchmarking tool,
  an Intel(R) Core(TM) i5-2500 CPU @ 3.30GHz, and
  2x8 Gb DDR3 RAM.
We assumed real subject names from Alexa's top-1M\footnote{%
  \url{http://s3.amazonaws.com/alexa-static/top-1m.csv.zip},
  accessed 2018-08-05.
} and average-sized certificates of 1500~bytes\footnote{%
  \url{https://www.grahamedgecombe.com/blog/2016/12/22/compressing-x509-certificates}, accessed 2018-08-15.
}, where a batch of $n$ subject names refers to the $n$ most popular domains.
Notably 689,245 certificates is the largest batch observed by us in
Google's Icarus log between 2017-01-25 and 2018-08-05, corresponding to an STH
interarrival time of 27.1~hours. The median (average) batch size and STH
interarrival time were 22818 (23751) certificates and 60.1 (61.6) minutes.
Only two batches were larger than 132077 certificates. Considering that
Icarus is one of the logs that see largest loads,\footnote{%
  \url{https://sslmate.com/labs/ct_growth/}, accessed 2018-08-15.
} we can make non-optimistic conclusions regarding the performance overhead of
LWM without inspecting other logs.

\begin{figure}
  \centering
  \includegraphics[width=8cm]{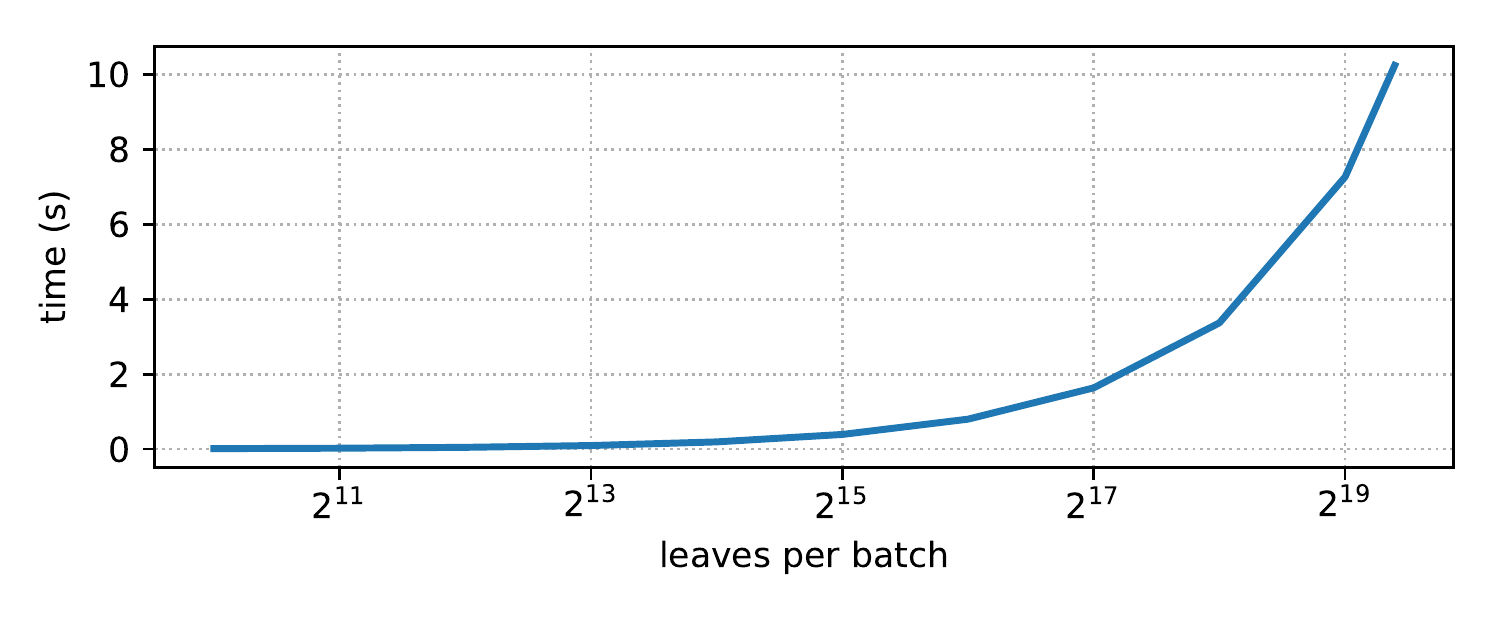}
  \caption{%
    Snapshot creation time as a function of batch size.
  }
  \label{fig:snapshot}
\end{figure}

Figure~\ref{fig:snapshot} shows snapshot creation time as a function of batch
size. Nearby the median ($2^{15}$) it takes 0.39~seconds to create a
snapshot from scratch, initializing state from an unordered dictionary and
caching all hashes for the first time. For the largest batch, the snapshot
creation time is roughly 10 seconds. Arguably this overhead is still
insignificant for logs, monitors, and notifiers because the associated STH
interarrival times are orders of magnitude larger.

\begin{figure}
  \centering
  \includegraphics[width=8cm]{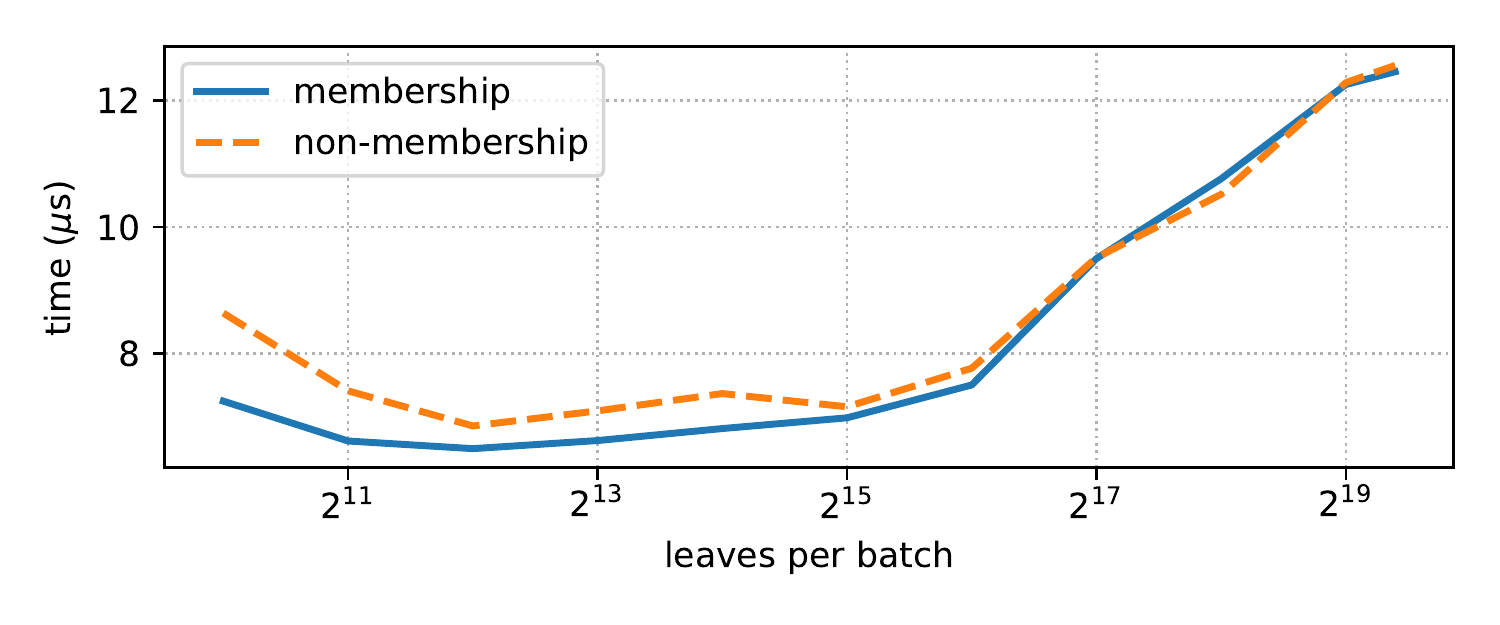}
  \caption{%
    Membership and non-membership proof query time as a function of batch
    size for a single and no match, respectively.
  }
  \label{fig:proofgen}
\end{figure}

\begin{figure}
  \centering
  \includegraphics[width=8cm]{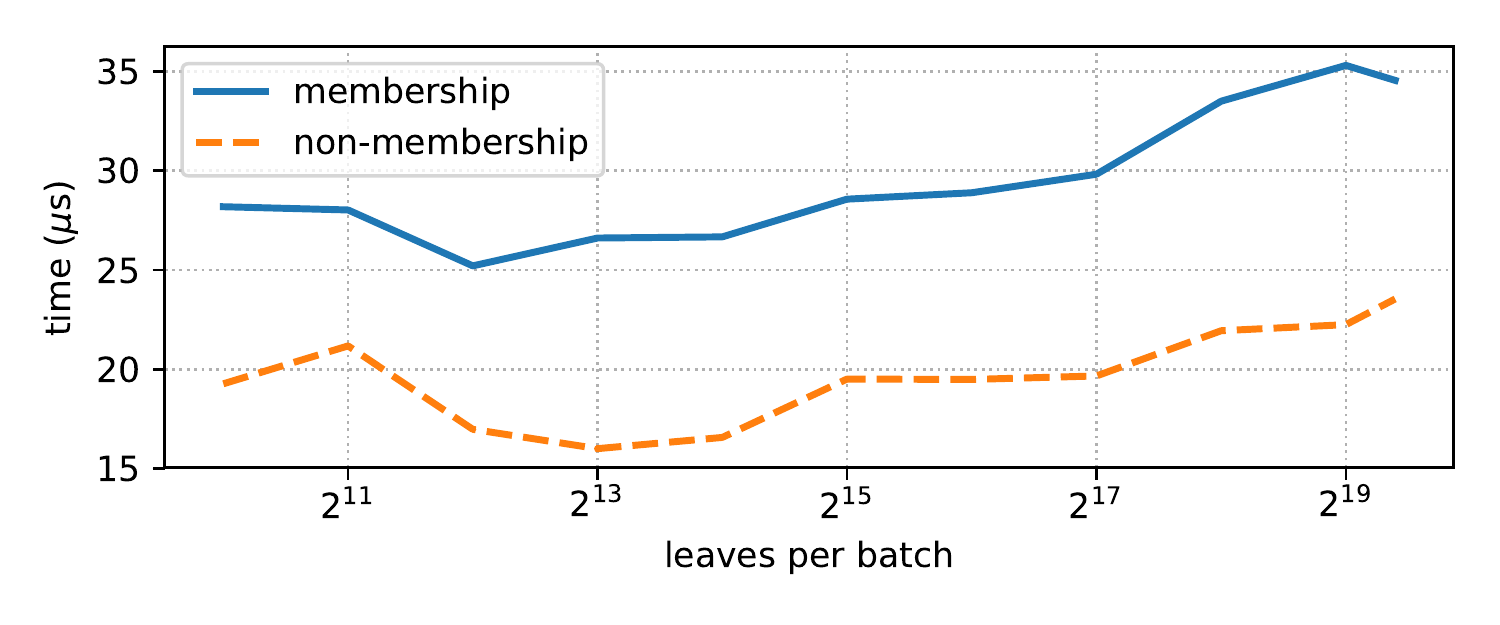}
  \caption{%
    Membership and non-membership verification time as a function of batch
    size for a single and no match, respectively.
  }
  \label{fig:proofvf}
\end{figure}

Figure~\ref{fig:proofgen} shows proof generation time as a function of batch size
while querying for the longest wild-card prefix with a single match
(membership), as well as another wild-card prefix without any match in com's
top-level domain (non-membership).
There is little or no difference between the generation time for these types
of wild-card proofs, and nearby the median it takes
around 7~$\mu s$. For the largest batch, this increased to $12.5$~$\mu s$.
A notifier can thus generate 288 million non-membership
notifications per hour \emph{on a single core}. Verification is also in the
order of $\mu s$, which should be negligible for a subject
(see Figure~\ref{fig:proofvf}).

\begin{figure}
  \centering
  \includegraphics[width=8cm]{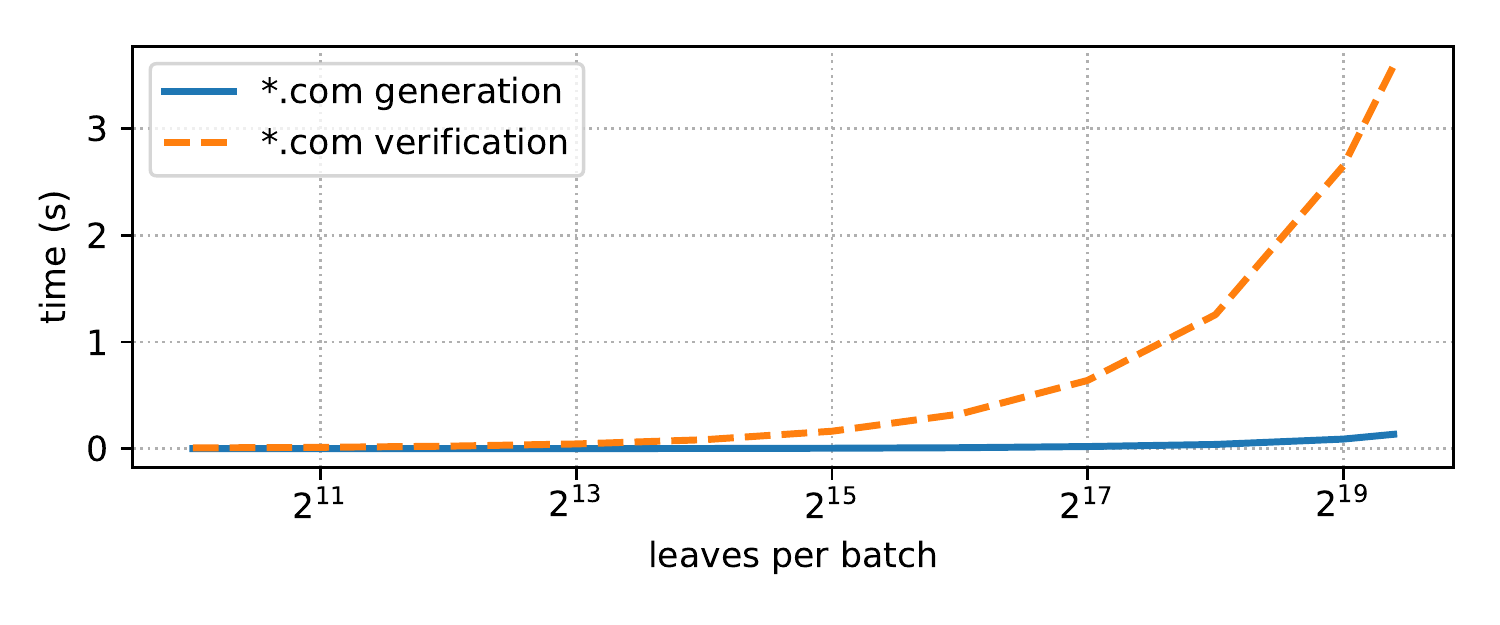}
  \caption{%
    Membership query and verification time for $\mathsf{*.com}$.
  }
  \label{fig:proofcom}
\end{figure}

To evaluate the cost of generating and verifying a wild-card notification with
a large number of matches, we queried for com's entire top-level domain
  (see Figure~\ref{fig:proofcom}).
In the largest batch where there are 352,383 matches, the proof
generation time is still relatively low: 134~ms. This corresponds to 28.9k
notifications per hour on a single core. The verification time is much larger:
3.5~seconds. This is expected since verification involves reconstructing the
root from all the matching leaves, which is at least as costly as creating a
snapshot of the same size
  (cf.\ $2^{18}$ in Figure~\ref{fig:snapshot}).
While these are relevant performance numbers, anyone who is interested in a
top-level domain would likely just download the entire batch.

Finally, the space \emph{overhead} of a verifiable wild-card notification is
dominated by the two audit paths that enclose the matching subject names. Given
that an audit path contains at most $\ceil{\log_2 n}$ sibling hashes for a batch
of size $n$, the median overhead is roughly one Kb per STH, log, and LWM
subject.  Viewed from the perspective of a self-monitor, this is a significant
bandwidth improvement:
  as opposed to downloading the median batch of 32.6~Mb,
  one Kb and any matching certificate(s) suffice.
In the case of multiple logs, the bandwidth improvement is even greater. For
the notifier we already established that it is relatively cheap to generate new
notifications. Namely, in the single-core case of 288~M notifications per hour
the bandwidth overhead would be 640~Mbs (i.e., all proofs must be distributed
before the next STH is issued). A notifier can thus notify for a dozen of logs
and a significant amount of LWM subjects without running into any CPU or
bandwidth restrictions. Notably this is under the assumption of a sound STH
frequency---one hour in our evaluation, as used by Icarus and many other logs.

\subsection{Related Work} \label{sec:eval:related}
Earlier work related to transparent certificate and key management often use
dynamic authenticated dictionaries~\cite{pad,accumulator,vds,aki}.
CONIKS maps a user's mail address to her public key in a binary Merkle prefix
tree, and after each update a client self-monitors her own key-binding by
fetching an exact-match (non-)membership proof~\cite{coniks}. While our work
is conceptually similar to CONIKS since a subject receives one (non-)membership
proof per log update, the main difference is that LWM builds a new Merkle tree
for each update in which wild-card queries are supported. This idea is
inapplicable for CONIKS because a user is potentially interested in the public
key of any mail address (hence the ability to query the entire data structure
on an exact-match).
CONIKS is similarly inapplicable for self-monitors in CT because a subject cares
about \emph{wild-card queries} and \emph{new certificates}.
Without the need for wild-cards, any authenticated dictionary could be used as
a batch building block to instantiate LWM.
While a radix tree viewed as a Merkle tree\footnote{%
  \url{https://github.com/ethereum/wiki/wiki/Patricia-Tree}, accessed 2018-08-15.
} could support efficient wild-card proofs, it is more complex than necessary.
Therefore, we built upon the work of Kocher~\cite{crt} and Nuckolls~%
\cite{range-mt} with a twist on how to group the data for a new use-case: LWM.

  \section{Conclusion} \label{sec:conclusion}
We proposed a backwards-compatible CT/bis extension that enables light-weight
monitoring (in short LWM). At the cost of a few hundred Kb per day, a subject
can either self-monitor or subscribe to verifiable certificate notifications for
a dozen of logs via an untrusted notifier. The security of LWM piggybacks on the
gossip-audit model of CT, and it relies only on the existence of at least one
honest monitor that verifies our extension. The cost of a compliant log is
overhead during the tree head construction, and this overhead is insignificant
in comparison to a log's STH frequency. A notifier can generate verifiable
certificate notifications---even for wild-card queries for all domains under a
top-level domain---in the order of milliseconds on a single core. Given an
STH frequency of one hour and 288~M LWM subjects, the incurred bandwidth
overhead is roughly 640~Mbps for proofs. As such, a log could easily be its
own notifier on a 1~Gbps connection. Further, any willing third-party could
notify for a dozen of logs on a 10~Gbps connection.

  \subsubsection{Acknowledgments.}
  We would like to thank Linus Nordberg for value feedback. This research was
funded by the Swedish Knowledge Foundation as part of the HITS research profile.

  \bibliographystyle{splncs04}
  \bibliography{src/references.bib}

\begin{thebibliography}{10}
\providecommand{\url}[1]{\texttt{#1}}
\providecommand{\urlprefix}{URL }
\providecommand{\doi}[1]{https://doi.org/#1}

\bibitem{sthnp}
Chuat, L., Szalachowski, P., Perrig, A., Laurie, B., Messeri, E.: Efficient
  gossip protocols for verifying the consistency of certificate logs. In:
  {IEEE} Conference on Communications and Network Security ({CNS}). pp.
  415--423 (Septemper 2015)

\bibitem{history-tree}
Crosby, S.A., Wallach, D.S.: Efficient data structures for tamper-evident
  logging. In: 18th {USENIX} Security Symposium. pp. 317--334 (August 2009)

\bibitem{pad}
Crosby, S.A., Wallach, D.S.: Authenticated dictionaries: {Real}-world costs and
  trade-offs. {ACM} Transactions on Information and System Security ({TISSEC})
  \textbf{14}(2),  17:1--17:30 (September 2011)

\bibitem{ctga}
Dahlberg, R., Pulls, T., Vestin, J., H{\o}iland{-}J{\o}rgensen, T., Kassler,
  A.: Aggregation-based gossip for certificate transparency. CoRR
  abs/1806.08817  (August 2018)

\bibitem{accumulator}
Derler, D., Hanser, C., Slamanig, D.: Revisiting cryptographic accumulators,
  additional properties and relations to other primitives. In: Topics in
  Cryptology---Proceedings of the Cryptographer's Track at the {RSA} Conference
  ({CT-RSA}). pp. 127--144 (April 2015)

\bibitem{ca-ecosystem}
Durumeric, Z., Kasten, J., Bailey, M., Halderman, J.A.: Analysis of the {HTTPS}
  certificate ecosystem. In: Proceedings of the 2013 Internet Measurement
  Conference. pp. 291--304 (October 2013)

\bibitem{vds}
Eijdenberg, A., Laurie, B., Cutter, A.: Verifiable data structures. Google
  research document (November 2015),
  \url{https://github.com/google/trillian/blob/master/docs/VerifiableDataStructures.pdf},
  accessed 2018-09-16

\bibitem{enisa}
ENISA: Certificate authorities---the weak link of {Internet} security. Info
  notes (September 2016),
  \url{https://web.archive.org/web/20180527220047/https://www.enisa.europa.eu/publications/info-notes/certificate-authorities-the-weak-link-of-internet-security},
  accessed 2018-09-16

\bibitem{katz}
Katz, J.: Analysis of a proposed hash-based signature standard. In: Third
  International Conference on Security Standardisation Research ({SSR}). pp.
  261--273 (December 2016)

\bibitem{aki}
Kim, T.H., Huang, L., Perrig, A., Jackson, C., Gligor, V.D.: Accountable key
  infrastructure {(AKI):} {A} proposal for a public-key validation
  infrastructure. In: 22nd International World Wide Web Conference ({WWW}). pp.
  679--690 (May 2013)

\bibitem{crt}
Kocher, P.C.: On certificate revocation and validation. In: Proceedings of the
  Second International Conference on Financial Cryptography ({FC}). pp.
  172--177 (February 1998)

\bibitem{ct}
Laurie, B., Langley, A., Kasper, E.: Certificate transparency. RFC~6962, IETF
  (June 2013), \url{https://tools.ietf.org/html/rfc6962}

\bibitem{ct/bis}
Laurie, B., Langley, A., Kasper, E., Messeri, E., Stradling, R.: Certificate
  transparency version 2.0. Internet-draft draft-ietf-trans-rfc6962-bis-28,
  IETF (March 2018),
  \url{https://tools.ietf.org/html/draft-ietf-trans-rfc6962-bis-28}, work in
  progress

\bibitem{coniks}
Melara, M.S., Blankstein, A., Bonneau, J., Felten, E.W., Freedman, M.J.:
  {CONIKS:} {Bringing} key transparency to end users. In: 24th {USENIX}
  Security Symposium. pp. 383--398 (August 2015)

\bibitem{mt}
Merkle, R.C.: A digital signature based on a conventional encryption function.
  In: Advances in Cryptology ({CRYPTO}). pp. 369--378 (August 1987)

\bibitem{ietf-gossip}
Nordberg, L., Gillmor, D.K., Ritter, T.: Gossiping in {CT}. Internet-draft
  draft-ietf-trans-gossip-05, IETF (January 2018),
  \url{https://tools.ietf.org/html/draft-ietf-trans-gossip-05}, work in
  progress

\bibitem{range-mt}
Nuckolls, G.: Verified query results from hybrid authentication trees. In:
  Proceedings of the 19th Annual {IFIP} {WG} 11.3 Working Conference on Data
  and Applications Security. pp. 84--98 (August 2005)

\bibitem{sth-push}
Sleevi, R., Messeri, E.: Certificate transparency in {Chrome}: Monitoring {CT}
  logs consistency. Design document, Google Inc. (March 2017),
  \url{https://docs.google.com/document/d/1FP5J5Sfsg0OR9P4YT0q1dM02iavhi8ix1mZlZe_z-ls/edit?pref=2&pli=1},
  accessed 2018-09-16

\bibitem{cosi}
Syta, E., Tamas, I., Visher, D., Wolinsky, D.I., Jovanovic, P., Gasser, L.,
  Gailly, N., Khoffi, I., Ford, B.: Keeping authorities ``honest or bust'' with
  decentralized witness cosigning. In: {IEEE} Symposium on Security and Privacy
  ({SP}). pp. 526--545 (May 2016)

\bibitem{ads}
Tamassia, R.: Authenticated data structures. In: 11th Annual European Symposium
  ({ESA}) on Algorithms. pp.~2--5 (September 2003)

\end{thebibliography}
\end{document}